\documentclass[twocolumn]{aa}
\usepackage{graphicx}
\usepackage{txfonts}
\usepackage{natbib}
\newcommand{\beq}   {\begin{equation}}
\newcommand{\eeq}   {\end{equation}}
\newcommand{\kms}   {km~s$^{-1}$}
\newcommand{\water}   {H$_2$O~}

\begin{document}
   \title{The magnetic field around late-type stars revealed by the circumstellar H$_2$O masers}

   \titlerunning{Magnetic Fields around Late-type stars}

%   \subtitle{}

   \author{W.H.T. Vlemmings\inst{1}\and
           H.J. van Langevelde\inst{2}\and
           P.J. Diamond\inst{3}
          }

   \offprints{WV (wouter@jb.man.ac.uk)}

   \institute{Department of Astronomy, Cornell University, Ithaca, NY 14853
         \and
         Joint Institute for VLBI in Europe, Postbus 2, 
                7990~AA Dwingeloo, The Netherlands
         \and
         Jodrell Bank Observatory, University of Manchester, Macclesfield,
                    Cheshire, SK11 9DL, England     
                          }

   \date{16-01-2005}

   \abstract{ Through polarization observations, circumstellar masers
are excellent probes of the magnetic field in the envelopes of
late-type stars. Whereas observations of the polarization of the SiO
masers close to the star and on the OH masers much further out were
fairly commonplace, observations of the magnetic field strength in the
intermediate density and temperature region where the 22~GHz \water
masers occur have only recently become possible. Here we present the
analysis of the circular polarization, due to Zeeman splitting, of the
\water masers around the Mira variable stars U~Her and U~Ori and the
supergiant VX~Sgr. We present an upper limit of the field around U~Her
that is lower but consistent with previous measurements, reflecting
possible changes in the circumstellar envelope. The field strengths
around U~Ori and VX~Sgr are shown to be of the order of several
Gauss. Moreover, we show for the first time that large scale magnetic
fields permeate the circumstellar envelopes of an evolved star; the
polarization of the \water masers around VX~Sgr reveals a dipole field
structure. We discuss the coupling of the magnetic field with the
stellar outflow, as such fields could possibly be the cause of
distinctly aspherical mass-loss.   }

   \maketitle

\section{Introduction}

At the end of their evolution, a large majority of stars go through a
period of high mass loss while climbing the asymptotic giant branch
(AGB). This mass loss, of the order of $10^{-7}$ to $10^{-4}$
M$_\odot$yr$^{-1}$, is responsible for the formation of a
circumstellar envelope (CSE). It is also the main source for
replenishing interstellar space with processed material.  AGB stars
such as the Mira variable stars, with main sequence masses less than a
few M$_\odot$, will eventually develop as planetary nebulae (PNe). The
heavier evolved stars such as the supergiants will eventually explode
as a supernova (SN). The exact role of magnetic fields in the mass
loss mechanism and the formation of CSEs is still unclear but could be
considerable. The study of several maser species found in CSEs has
revealed important information about the strength and structure of
magnetic fields throughout the envelopes surrounding the late-type
stars. At distances from the central star of up to several thousands
of AU, measurements of the Zeeman effect on OH masers indicate
magnetic fields strengths of a few milliGauss \citep[e.g.][]{SC97,
MvL99}. Additionally, weak alignment with the CSE structure is found
\citep[e.g.][]{E04}. Observations of SiO maser polarization have shown
highly ordered magnetic fields close to the central star, at radii of
5-10 AU where the SiO maser emission occurs
\citep[e.g.][]{BM87,KD97}. When interpreting the circular polarization
of the SiO masers as standard Zeeman splitting, the magnetic field
strength determined from these observations could be up to several
tens of Gauss. However, a non-Zeeman interpretation can explain the
observations by magnetic field strengths of only several tens of
milliGauss \citep{WW98}. 
%
%{\bf A recently reported correlation between the linear and circular
%polarization of SiO masers \citep{H03} might be the signature of this
%non-Zeeman polarization mechanism. However, the observations of
%the circular polarization of H$_2$O masers paint a different picture
%\citep[][hereafter V02]{V02b}.}
%

 Recently, high circular polarization of circumstellar H$_2$O
masers was observed for a small sample of late-type stars
\citep[][hereafter V02]{V02b}.
H$_2$O masers occur at intermediate distances in the CSE, in gas that
is a factor of $10-1000$ more dense than the gas in which OH masers
occur. \citet[][hereafter V01]{V01} managed to detect the Zeeman
splitting of the circumstellar H$_2$O maser around the supergiant star
S~Per, even though for typical field strengths of a few hundred mG the
H$_2$O Zeeman splitting is extremely small, only $\approx 10^{-3}$
times the typical half-power width of the H$_2$O maser line
($\Delta\nu_L \approx 30$~kHz). Additional detections were discussed
in V02 and typical field strength of between $0.2$ and $1$~G were
found, using the full, non-LTE radiative transfer equations from
\citet{NW92}. 
%Such high magnetic field strengths at a few hundred AU
%from the central star indicate that the standard Zeeman interpretation
%of the SiO maser polarization is most likely correct.
 Such high magnetic field strengths at a few hundred AU from the
central star give strong support for the standard Zeeman
interpretation of the SiO maser polarization.

The results from V02 indicate that the magnetic field strengths
measured in the circumstellar maser regions are consistent with either
a $r^{-2}$ field strength dependence on the increasing distance to the
star, similar to a solar-type magnetic field or possibly an $r^{-3}$
dependence as produced by a dipole magnetic field. This implies
surface magnetic fields of hundreds to several thousand Gauss,
indicating that the magnetic field pressure dominates the thermal and
radiation pressure at the surface of the star. As a result, the
magnetic field likely plays a very important role in driving the
late-type star mass loss. Additionally, magnetic fields possibly play
an important role in shaping the CSEs. This will also affect the
formation of PNe, that are often observed to have distinctly
non-spherical shapes.

 Here we present new results on the magnetic field strength in the
envelopes of several late-type stars. We discuss the observations and
data calibration in \S~\ref{obs} and the analysis method in
\S~\ref{method}. The new and previous results for our sources are
presented in \S~\ref{results}. In \S~\ref{vxsgr} the specific 
results on the supergiant VX~Sgr are discussed, where the maser polarization
observations allow for the investigation of the morphology of the
magnetic field. The consequences of these measurements are discussed
in \S~\ref{disc} and are followed by our conclusions in
\S~\ref{concl}.

\section{Observations}
\label{obs}

The observations were performed at the NRAO Very Long Baseline Array
(VLBA) on April 20 2003. The average beam width is $\approx 0.5 \times
0.5$~mas at the frequency of the $6_{16} - 5_{23}$ rotational
transition of H$_2$O, 22.235 GHz. We used 4 baseband filters of 1 MHz
width, which were overlapped to get a velocity coverage of $\approx
44$~km/s, covering most of the velocity range of the H$_2$O
masers. Similar to the observations in V02, the data were correlated
multiple times. The initial correlation was performed with modest
($7.8$~kHz$ = 0.1$~\kms) spectral resolution, which enabled us to
generate all 4 polarization combinations (RR, LL, RL and LR). Two
additional correlator runs were performed with high spectral
resolution ($1.95$~kHz$ = 0.027$~\kms) which therefore only contained
the two polarization combinations RR and LL. This was necessary to be
able to detect the signature of the H$_2$O Zeeman splitting in the
circular polarization data and to cover the entire velocity range of
the \water masers. Each source-calibrator pair was observed for 6
hours. The calibrator was observed for $1.5$ hours in a number of
scans equally distributed over the 6 hours.

\subsection{Calibration}

 The data analysis path is described in detail in V02. It follows the
 method of \citet{KDC95} and was performed in the
 Astronomical Image Processing Software package (AIPS). The
 calibration steps were performed on the data-set with modest spectral
 resolution. We were also forced to flag several channels that
 suffered from strong interference. Fringe fitting and
 self-calibration were performed on a strong maser feature. The
 calibration solutions were then copied and applied to the high
 spectral resolution data-set.  Finally, corrections were made for
 instrumental feed polarization using a range of frequency channels on
 the maser source, in which the expected frequency averaged linear
 polarization is close to zero. Then image cubes were created for
 stokes I,Q,U and V in the modest spectral resolution data set, and
 for stokes I and V in the high spectral resolution data-set.

\subsection{Sources}

\begin{table*}
\caption{Star Sample}
\begin{tabular}{|l|c|cc|c|c|c|c|}
\hline
Name & Type & RA (J2000) & Dec (J2000) & Distance & Period & V$_{\rm rad}$ & Peak
flux\\
&&($^{h}~^{m}~^{s}$)&($^{\circ}~{'}~{"}$)&(pc)&(days)&(km/s)&(Jy)\\
\hline
\hline
VX~Sgr & Supergiant &  18 08 04.0485 & -22 13 26.614 & 1700$^a$ & 732 & 5.3 & 43.2\\
U Her & Mira & 16 25 47.4713 & +18 53 32.867 & 277$^b$ & 406 & -14.5 & 9.75
\\
U~Ori & Mira & 05 55 49.1689 & +20 10 30.687 & 300$^c$ & 368 &
-38.1 & 6.44\\
R~Cas & Mira & 23 58 24.8725 & +51 23 19.703 & 176$^b$ & 430 &
26.0 & -\\
\hline
\multicolumn{8}{l}{$^a$ \citet{M96},$^b$\citet{V03}, $^c$\citet{KY98}}\\
\end{tabular}
\label{sample}
\end{table*}

We observed 4 late-type stars, the supergiant VX Sgr and the Mira
variable stars U~Ori, R~Cas and U~Her. U~Her was previously observed
in December 1998 as part of the observations described in
V02. Unfortunately, we were unable to detect the H$_2$O masers of
R~Cas, presumably due to source variability. In Table~\ref{sample} we
list the observed sources with type, position, distance, period and
velocity, as well as the peak \water maser flux at the epoch of
observation. In the high spectral resolution total intensity channel
maps, the noise is dominated by dynamic range effects and is $\approx
60$~mJy. In the circular polarization polarization maps the rms noise
is $\approx 20$~mJy.

The polarization of the OH masers of our sources has previously been
observed and SiO maser polarization measurements have been performed
on VX~Sgr. Polarization of the 1612 MHz OH masers around U~Ori
indicate a magnetic field of $B_{\rm OH}\approx 10$~mG \citep{R79},
while \citet{PF00} found a field of $B_{\rm OH}\approx 1$~mG on the
1665 and 1667 MHz OH masers around U~Her. The magnetic fields in the
main-line and satellite-line OH maser regions around VX~Sgr were
measured using MERLIN \citep{CC86, T98, SC01} and the VLA
\citep{ZF96}. The polarization of the 1665 and 1667~MHz OH masers
indicated a field strength of $B_{\rm OH}\approx 2$~mG, while the
field in the 1612 MHz maser region is $B_{\rm OH}\approx 1$~mG. In the
single dish observations of the circular polarization of SiO masers by
\citet{BM87}, they observe a circular polarization
percentage of 8.7\%. This indicates $B_{\rm SiO}\approx 90$~G for
VX~Sgr.

\section{Method}
\label{method}

For the analysis of the polarization spectra we used both the basic
LTE interpretation and the full radiative transfer non-LTE
interpretation, which were thoroughly described in V02. The LTE
analysis consists of a standard Zeeman interpretation assuming LTE. As
a result, the narrowing and rebroadening of the maser profile caused
by saturation are not reproduced. We have included the
possibility of multiple masing hyperfine lines. In this analysis, the
magnetic field strengths are determined by fitting a synthetic
V-spectrum, proportional to $dI/d\nu$, to the polarization
spectra. The magnetic field strengths follow from

\begin{eqnarray}
%\begin{equation}
P_{\rm V} & = & (V_{\rm max} - V_{\rm min})/I_{\rm max} \nonumber\\
& = & 2\cdot A_{\rm F-F'}\cdot B_{\rm [Gauss]} \rm{cos}\theta/\Delta v_{\rm L}[\rm{km~s}^{-1}]. 
\label{eq2}
%\end{equation}
\end{eqnarray}
with $P_V$ the percentage of circular polarization, $V_{\rm max}$ and
$V_{\rm min}$ the maximum and minimum of the synthetic LTE V-spectrum
fitted to the observations. $I_{\rm max}$ and $\Delta v_{\rm L}$ are
the peak flux and the full width half-maximum (FWHM) of the maser
feature respectively. $B$ is the magnetic field strength at an angle
$\theta$ from the line of sight. The $A_{\rm F-F'}$ coefficient
depends on the masing hyperfine component and in V02 was estimated to
be $\approx 0.011$, which is used here for all LTE estimates.

For the non-LTE analysis, the coupled equations of state for the 99
magnetic substates of the three dominant hyperfine components from
\citet{NW92} were solved for a linear maser in the
presence of a magnetic field. The modeled spectra were then directly
fitted to the observed spectra. When a direct model fit was not
possible, we used Eq.\ref{eq2} with $A_{\rm F-F'} = 0.018$, which is
the corresponding best estimate for the coefficient in the non-LTE
case.

The spectral fitting for both the non-LTE as well as the LTE analysis
requires the removal of the scaled down total power spectrum from the
V-spectrum to correct for small residual gain errors between the
right- and left-polarized antenna feeds.  It was found in V01
and in V02 that generally, the V-spectra were narrower than expected
in the LTE case. This could be partly explained by introducing the
non-LTE method, however, some narrowing remained. In V02, we attributed
this to the application of uni-directional maser propagation in the
modeling, and we found that a bi-directional maser could 
explain some of the narrowing. However, for computational
efficiency we continue to use the uni-directional method, while
allowing for additional narrowing of the V-spectrum, as we found that
the estimates for the fractional circular polarization are robust.

\section{Results}
\label{results}

\begin{table*}
\caption{Results}
\begin{tabular}{|l|c|c|c|c|c|c|c|}
\hline
Name & Feature  & Flux (I) & $V_{\rm rad}$ & $\Delta
v_{\rm L}$ & P$_{\rm V}$ &
B~${\rm cos}\theta$ LTE & B~${\rm cos}\theta$ non-LTE \\
 & & (Jy) & ${\rm (km/s)}$ & ${\rm (km/s)}$ & $(\times10^{-3})$ &
${\rm (mG)}$& ${\rm (mG)}$\\
\hline
\hline
VX Sgr & a & 18.8 & -2.01 & 0.47 & 21.0 & 350$\pm$72 & 512$\pm$90 \\  
 & b$^*$ & 11.7 & -1.28 & 0.38 & 22.6 & -307$\pm$99 & -248$\pm$62 \\
 & c & 6.9 & 7.18 & 0.61 & 171.2 & 3705$\pm$249 & 3375$\pm$265 \\
 & d & 23.1 & 7.16 & 0.52 & 32.5 & -609$\pm$67 & -469$\pm$60 \\
 & e & 18.9 & 6.81 & 0.53 & 19.2 & 359$\pm$79 & 249$\pm$46 \\
 & f & 36.7 & 16.43 & 0.57 & & & $<$ 54 \\  
 & g & 43.2 & 13.77 & 0.58 & 3.5 & -75$\pm$34 & -63$\pm$20 \\
 & h & 9.6 & 11.86 & 0.57 & 15.5 & -326$\pm$184 & -290$\pm$115 \\
 & i & 23.3 & 12.73 & 0.48 & & & $<$ 99 \\  
 & j & 9.4 & -0.9 & 0.62 & 16.4 & -370$\pm$202 & -325$\pm$126 \\
 & k & 6.3 & -0.7 & 0.56 & 14.4 & 285$\pm$281 & 277$\pm$195 \\
 & l & 14.9 & 6.73 & 0.51 & 66.7 & 1231$\pm$123 & 966$\pm$120 \\
 & m & 9.7 & 20.6 & 0.52 & 9.9 & -188$\pm$140 & -167$\pm$87 \\
 & n$^{**}$ & 14.5 & -0.1 & 0.87 & 162.7 & -4954$\pm$238 & -4082$\pm$305 \\
 & o & 6.8 & 21.19 & 0.71 & & & $<$ 381 \\ 
 & p & 6.3 & 22.17 & 0.62 & & & $<$ 471 \\
 & q & 8.5 & 20.97 & 0.49 & & & $<$ 239 \\
 & r & 3.4 & 20.93 & 0.62 & & & $<$ 687 \\
 & s & 2.7 & 9.95 & 0.54 & & & $<$ 987 \\
 & t & 8.4 & 16.26 & 0.48 & & & $<$ 217 \\
 & u & 9.9 & 13.64 & 0.58 & 4.5 & -94$\pm$154 & -85$\pm$99 \\
 & v & 7.2 & 7.1 & 0.51 & 39.9 & -738$\pm$194 & -366$\pm$75 \\
 & w$^{**}$ & 4.7 & 1.17 & 0.56 & 109.6 & 2160$\pm$387 & 2195$\pm$251 \\
 & x & 5.1 & 18.24 & 0.69 & & & $<$ 512 \\ 
 & y & 6.8 & 2.52 & 0.46 & 61.1 & 1012$\pm$281 & 921$\pm$208 \\
\hline
U Her & a & 9.75 & -14.56 & 0.55 & & & $<$ 127 \\
 & b & 3.75 & -14.42 & 0.58 & & & $<$ 418 \\
 & c & 2.09 & -15.82 & 0.55 & & & $<$ 610 \\
\hline
U Ori & a$^*$  & 6.44 & -41.05 & 0.61 & 22.6 & -517$\pm$152 & -401$\pm$96 \\
 & b & 1.87 & -41.23 & 0.75 & 217.9 & -5770$\pm$718 & -3192$\pm$425 \\
 & c & 0.95 & -40.85 & 0.66 & 140.4 &  3462$\pm$1210 &  1398$\pm$382 \\
\hline
\multicolumn{8}{l}{$^*$ no accurate fit possible (see text)} \\
\multicolumn{8}{l}{$^{**}$ suffer from interference (see text)}
\end{tabular}
\label{Table:results}
\end{table*}

 We have determined the magnetic field strength on the strongest maser
features in the band ($44$~km/s wide). The results of both the LTE and
non-LTE analysis are shown in Table~\ref{Table:results} for features
with peak fluxes down to $\sim 5\%$ of the brightest maser spot. As
seen in column 6, the percentage circular polarization is relatively
large and varies between 0.3\% and 20\% with the highest polarization
being detected for U~Ori. Magnetic field strengths obtained using the LTE
Zeeman method are given in column 7 while the results obtained using
the non-LTE models are shown in column 8. Features that did not show
circular polarization at a level higher than 3 times the rms noise in
the V-spectrum are considered non-detections. However, the formal
errors on the magnetic field strength also include the errors
on total intensity and line width, and are thus larger, resulting in
occasional magnetic field strengths around the $1\sigma$ noise level. For the
non-detections we have determined the {\it absolute} upper limits
using the $A_{\rm F-F'}$ coefficient obtained in the non-LTE radiative
transfer models. In the LTE analysis the limits would be larger by a
factor of $\approx 1.4$

Maser features that had circular polarization spectra that could not
be fitted accurately are labeled in Table.~\ref{Table:results}. This
is likely due to blending of several features, both spatially and
along the velocity axis. The values for the magnetic field strength
for those features were determined using only the minimum $V_{\rm
min}$ and maximum $V_{\rm max}$ of the V-spectrum after it was forced
to be point symmetric.  The features labeled $**$ (VX~Sgr~n, shown in
Fig.\ref{Fig:vxsgrspec} and VX~Sgr~w) suffered from external
interference, most of which we managed to remove. Some effects remain
however, and it is unclear if this had an effect on the measured
magnetic field strength for those features.

 As in V02, we were unable to detect any significant linear
polarization above a limit of $\sim 0.02\%$ for the strongest maser
features and a few \% for most of the weakest. There was a hint of
linear polarization on the strongest maser feature of U~Ori, but as
the level of polarization was at most only at the $1\sigma$ level of
$\approx 2\%$ it cannot unambigiously be classified as a detection.

\subsection{U Her}
\begin{figure}
%\epsscale{1.0} \plotone{uher2.eps}
   \resizebox{\hsize}{!}{\includegraphics{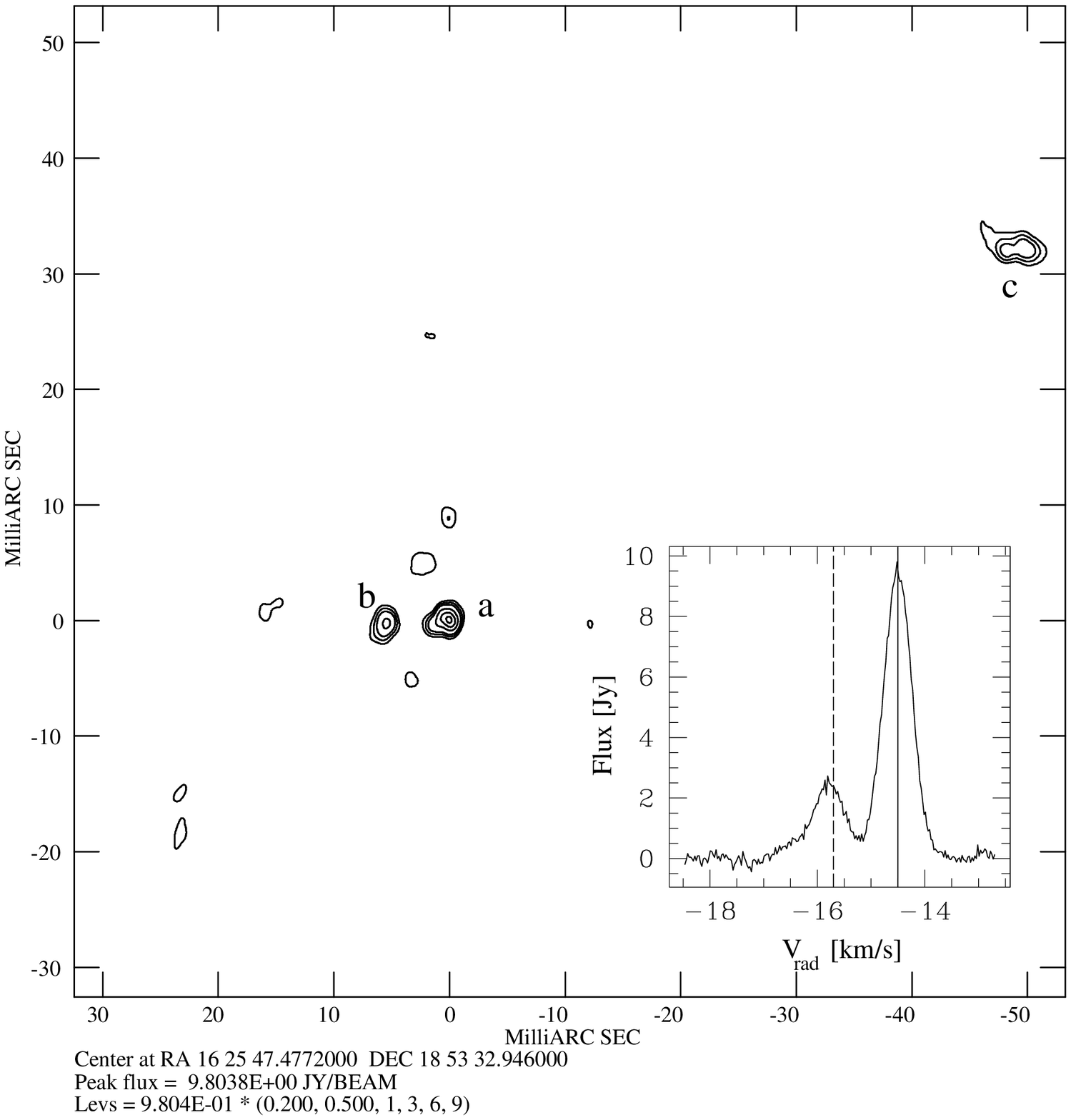}}
   \hfill
\caption[U Her]{Total intensity image of the H$_2$O maser features
around U~Her. The inlay shows the total intensity spectrum of the
U~Her \water masers, where the solid vertical line indicates the
stellar velocity and the dashed vertical line indicates the velocity
of the maser feature that was found to be coincident with stellar position
in \citet{V02a}.}
\label{Fig:uher}
\end{figure}

 After the observations of U~Her in V02 revealed a magnetic field in
the \water maser region of $\sim 1.5$~G, U~Her was observed again in
our second observational run. The integrated total intensity map of
the \water maser features is shown in Fig.~\ref{Fig:uher}. Even though
the strength of the U~Her maser features is similar in both
observational epochs, we did not detect any significant circular
polarization in the observations presented here. The upper limit we
find is $\sim 130$~mG on the strongest and $\sim 600$~mG on the
weakest maser feature. However, the spectrum and spatial distribution
of the masers has changed significantly in the more than 4 years
between the observational epoch of December 13th 1998 and April 20th
2003. The velocity of the maser features observed in V02 ($-19.3$ --
$-17.6$~km/s) was several km/s more blue-shifted than the maser
features observed here, that are close to the stellar radial velocity
of $-14.5$~km/s. Whereas the spatial distribution of the maser spots
in V02 showed an elongated structure in the N-S direction,
Fig.~\ref{Fig:uher} indicates a significantly different structure.
The weakest VLBI feature detected at $-15.82$~km/s (labeled $c$ in
Fig.~\ref{Fig:uher}) corresponds in velocity to the feature that was
found to be the strongest in the MERLIN observations of
\citet{V02a}. In those observations, this feature was found to be
aligned with the optical position to within $15$~mas and interpreted
as the amplified stellar image.
%to discussion ?
Assuming that the current feature at $-15.82$~km/s is also the
amplified stellar image we obtain a lower limit on the distance on
the sky between the brighter \water maser features and the star. For a
distance to U~Her of $277$~pc, the projected separation between the
star and the two brightest maser features is $\approx 16$~AU. Since
the velocity of these maser features is close to the stellar velocity,
we expect the true separation to be similar. Assuming that the maximum
magnetic field strength in V02 was measured at the inner edge of the
\water maser region, which was found to be $\approx 12.5$~AU
\citep{BC03}, the expected magnetic field strength at $16$~AU is
$\approx 700$~mG assuming a dipole magnetic field configuration. The
difference between the expected field strength and the measured upper
limits could be due to a large angle $\theta > 75^\circ$ between the
magnetic field lines and the line of sight. Alternatively, some of the
magnetic field strength difference could be due to different maser
clump density by up to a factor of $\sim 10$, if the magnetic field lines
are frozen into the dense clumps. We could also be observing actual
variability in the magnetic field.

\subsection{U Ori}
\begin{figure*}[htf]
%\epsscale{1.0} \plotone{uori.eps}
   \resizebox{\hsize}{!}{\includegraphics{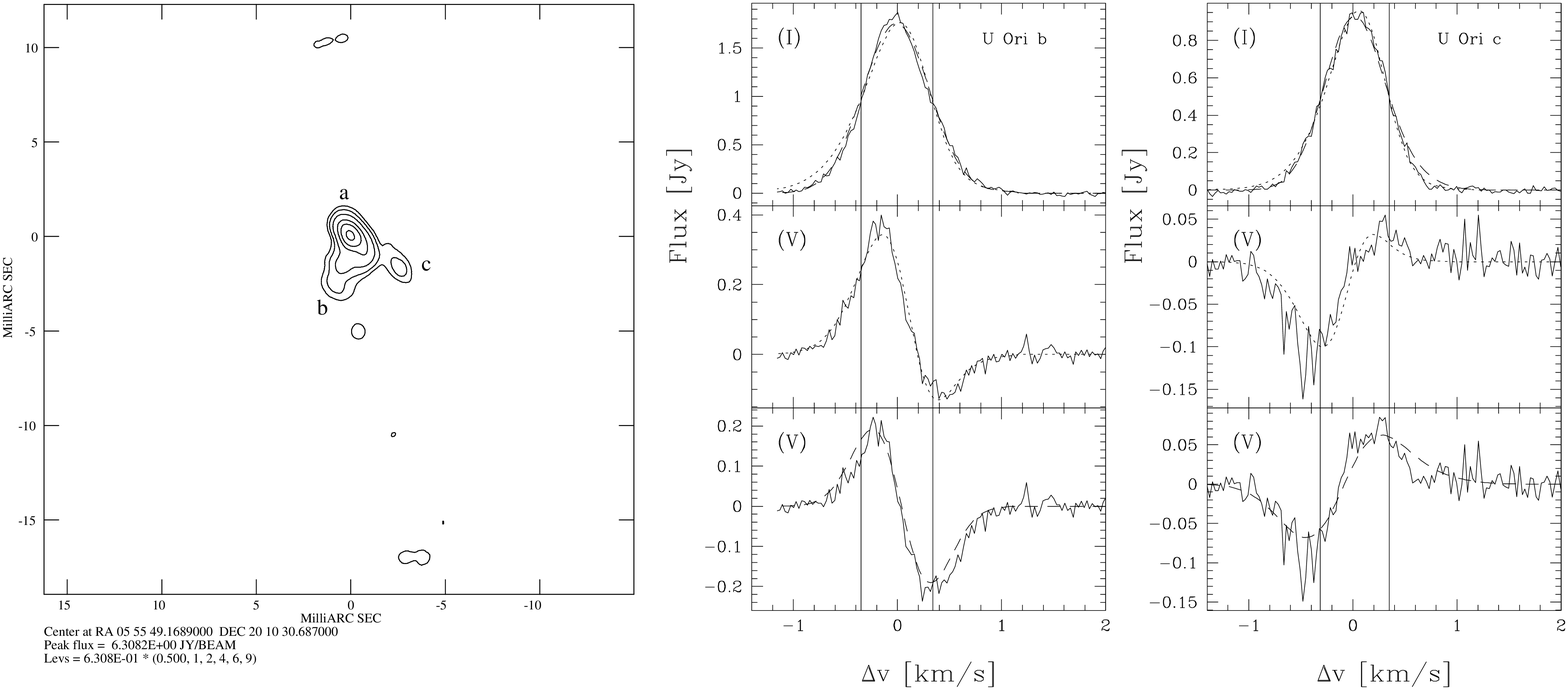}} \hfill
\caption[uori]{(left) Total intensity image of the H$_2$O maser
features around U~Ori. (right) Total power (I) and V-spectra for
selected maser features of U~Ori. The bottom panel shows the best
fitting synthetic V-spectrum produced by the standard LTE Zeeman
interpretation (dashed line). The middle panel shows the best non-LTE
model fit (dotted line). The corresponding total power fits are shown
in the top panel. The V-spectra in the lower two panels are adjusted
by removing a scaled down version of the total power spectrum as
indicated in \S~\ref{method}, which is different for the LTE and
non-LTE fits. The solid vertical lines show the expected position of
the minimum and maximum of the V-spectrum in the general LTE
interpretation.}
\label{Fig:uori}
\end{figure*}

Fig.~\ref{Fig:uori} shows the integrated total intensity \water maser map of
U~Ori. Very little structure is observed and the maser features seem
to be somewhat blended both spatially as well as in velocity. Strong
circular polarization up to $\approx 22\%$ is detected on the three
bright central features where interestingly the features $b$ and $c$,
to the left and right of the brightest feature $a$, show an opposite sign for
the magnetic field. The total intensity and circular polarization
spectra for these features are also shown in Fig.~\ref{Fig:uori}. From
these we estimate the absolute magnetic field strength in the
maser region to be $\sim 3.5$~G. While the central feature $a$ does
not show such a high field strength, the feature seems to be a blend
of two features with oppositely directed magnetic fields. The more
complex circular polarization spectrum of U~Ori $a$ can be explained
by two features with a magnetic field of $\sim 800$~mG of opposite
sign. The high magnetic field strength measured for the Mira variable
star U~Ori is consistent with the high field ($10$~mG) measured using the
OH masers \citep{R79} assuming a dipole $r^{-3}$ dependence of
the magnetic field on the distance to the star. This assumes that the
\water masers occur in a shell with an outer edge of $30$~AU, as
determined by \citet{BC03}.

\subsection{VX Sgr}
\begin{figure*}[htf]
%\epsscale{1.0} \plotone{vxsgr.eps}
   \resizebox{\hsize}{!}{\includegraphics{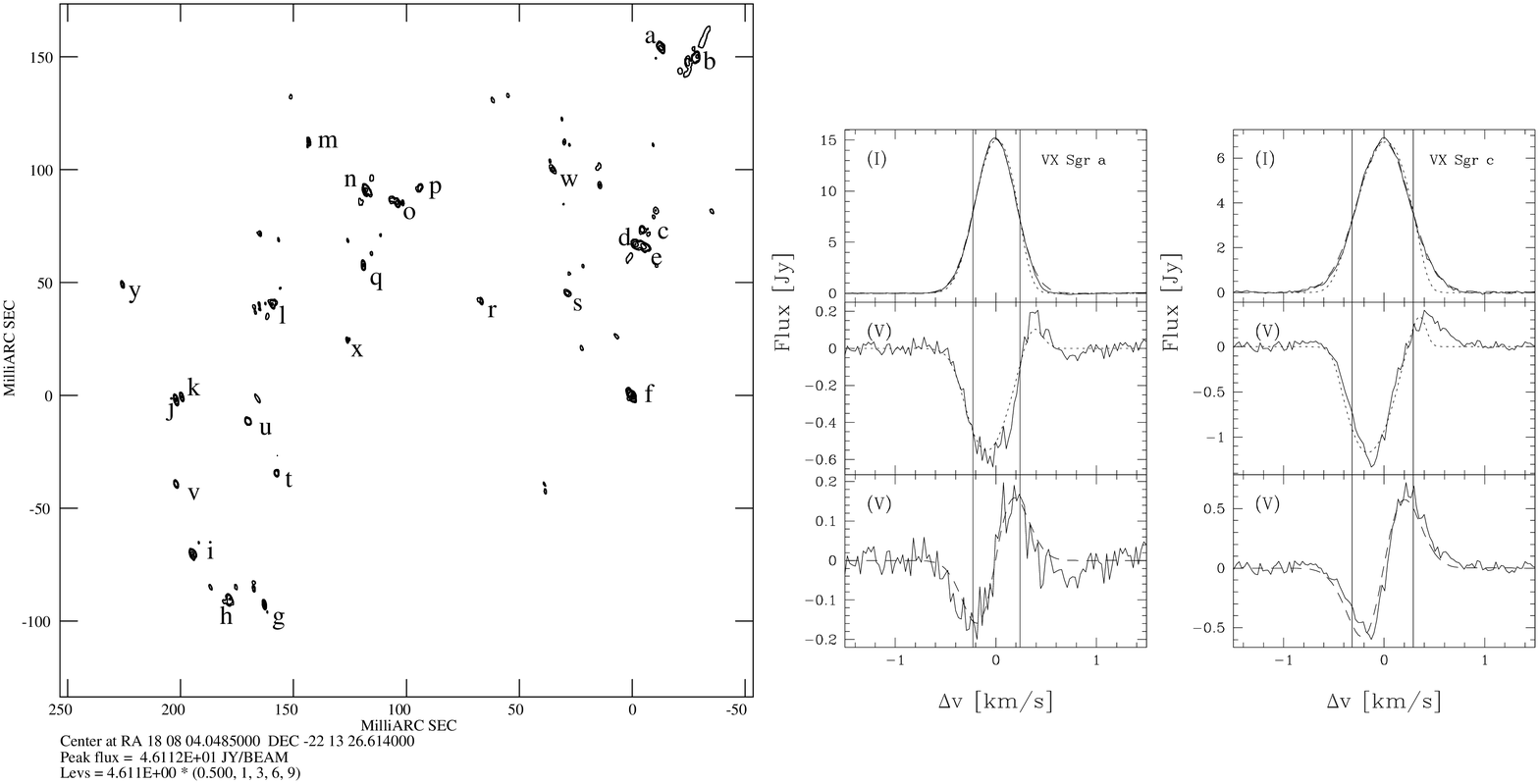}} \hfill
\caption[VX Sgr]{Similar to Fig.~\ref{Fig:uori} for VX~Sgr.}
\label{Fig:vxsgr}
\end{figure*}

\begin{figure*}[htf]
%\epsscale{1.0} \plotone{vxsgrspec.eps}
   \resizebox{\hsize}{!}{\includegraphics{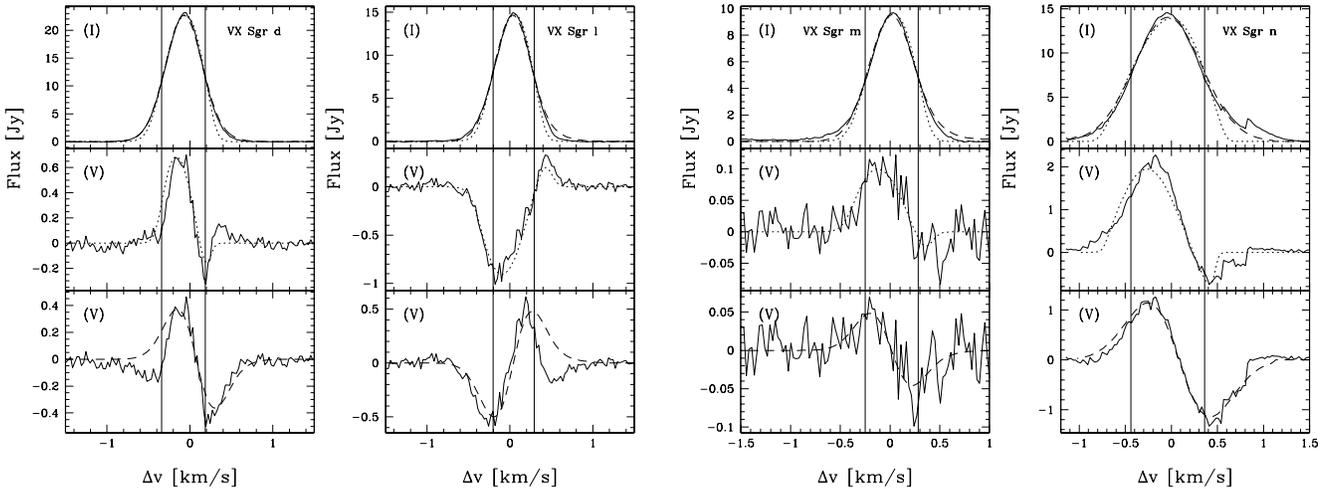}} \hfill
\caption[VX Sgr spec]{Similar to the spectra in Fig.~\ref{Fig:uori}
and Fig.~\ref{Fig:vxsgr} for additional maser features around
VX~Sgr. Feature VX~Sgr $n$ suffers from some interference effect in
the right wing of the spectrum but this does not affect the magnetic
field determination.}
\label{Fig:vxsgrspec}
\end{figure*}

The integrated total intensity \water maser map of the supergiant
VX~Sgr is shown in Fig.~\ref{Fig:vxsgr}. VX~Sgr shows a rich structure
of strong \water maser features in a $350\times250$~mas ellipse,
corresponding to $600\times425$~AU at a distance of $1.7$~kpc. The
maser feature distribution is similar to that seen in the MERLIN
observation by \citet[][hereafter M03]{MYRC03}. Circular polarization
was detected ranging from $0.3$--$16\%$ for a large number of maser features
as seen in Table~\ref{Table:results}. Several of the spectra are shown
in Figs.~\ref{Fig:vxsgr} and ~\ref{Fig:vxsgrspec}. As mentioned above,
a few of the maser features suffered from blending or external
interference. The maximum magnetic field strength measured is $\sim
4$~G. Across the maser features we find a clear transition between a
negative magnetic field in the S-E to a positive magnetic field in the
N-W. This is the first detection of a large scale magnetic field in a
circumstellar \water maser region.
%and the possible morphology of the
%field is discussed further in \S~\ref{vxsgr}.

\section{The Morphology of the Magnetic Field around VX Sgr}
\label{vxsgr}

 The maser shell of VX~Sgr has been studied extensively with MERLIN
and VLA observations \citep[e.g.][M03]{L84, CC86, ZF96, T98}. The recent
observations of M03 indicate that the \water masers arise in a thick
shell between $100$~AU and $325$~AU from the star. In this shell, the
\water masers seem to be accelerated from $\sim 10$~km/s to $20$~km/s.
The \water maser shell shows a clear elliptical asymmetry which was
modeled in M03 as a spheroidal maser distribution intersected with an
under-dense bi-conical region at an inclination angle $i=60^\circ \pm
30^\circ$ from the line of sight and rotated $\Theta = 200^\circ \pm
30^\circ$ on the plane of the sky. \citet{M96} also observed the
elliptical \water maser distribution with the VLBA and, constrained by
proper motions, fitted the emission region using an oblate spheroid
with a maximum radius of $250~$mas at an inclination angle $i\sim
46^\circ$ and projected position angle $\Theta \sim 230^\circ$.

 Observations of the 1612~MHz OH masers at $\sim 1400$~AU from the
star also show the elliptical asymmetry observed in the \water masers.
The magnetic field strength in the 1612~MHz OH maser region was
estimated to be of the order of a milliGauss \citep{T98}.
\citet{SC01} observed the linear polarization of the OH
masers and found structure that could be explained by a dipole
magnetic field with an inclination axis of about $20^\circ$ or
$30^\circ$. Additionally, the observations of Zeeman pairs in the OH
emission indicates a change in magnetic field direction across the
maser region.
% similar to the one we observed across the \water masers. 
This was suggested to be the signature of a dipole magnetic field
\citep{ZF96}. Further observations indicated that the magnetic
axis was projected at a position angle $\Theta = 210^\circ\pm20^\circ$
\citep{SC97}. Observations of the OH mainline masers
emission at 1667~MHz closer to the star are consistent with the same
magnetic angle \citep{RC00}.

\begin{figure*}[htf]
%\epsscale{1.0} \plotone{dipolebw2.eps}
   \resizebox{\hsize}{!}{\includegraphics{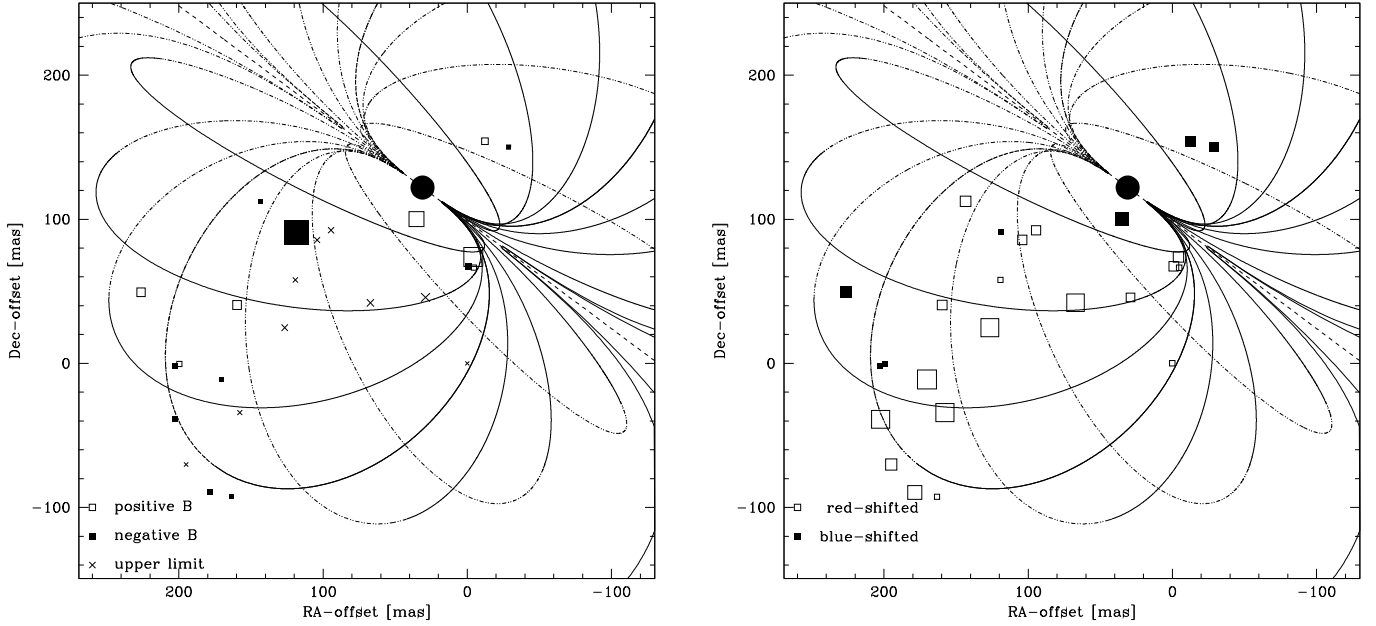}} \hfill
\caption[brplot]{The best fitted dipole magnetic field for the \water
maser observations around VX~Sgr (denoted by the solid circle). (left)
The distribution of the \water maser features indicating the measured
magnetic field strengths. Open symbols denote a positive magnetic
field while the closed symbols correspond to a negative magnetic
field. The crosses represent the upper limits. The symbols have been
scaled relative to the magnetic field strength. (right) The
distribution of the \water maser features indicating the velocity
structure of the maser features. The open symbols are the red-shifted
features and the solid symbols are the blue-shifted features. The size
indicates the velocity difference with the stellar velocity ($v_{\rm
rad} = 5.3$~\kms)}
\label{Fig:dipole}
\end{figure*}

We have fitted a dipole magnetic field to our \water maser magnetic
field observations of VX~Sgr. As the positions of the maser features
in the shell along the line of sight are unknown, we used the
accelerating spherical outflow model of \citet{CC86} to describe the
velocity structure of the maser shell. Thus, the observed radial
velocity of the maser features directly maps into the third spatial
dimension, allowing for a three dimensional fit. A maximum likelihood
fit was made to the dipole field, solving for the stellar position,
the inclination and position angle of the magnetic field and the field
strength at the surface of VX~Sgr. The stellar radius was fixed at
$\sim 16$~AU, in agreement with the observations by \citet{DB94} and
\citet{G95}. In the fits we also included the maser features for which
we only determined an {\it absolute} upper limit. The errors on the
magnetic field strength were taken from our analysis, while we
included errors of $1$~km/s on the velocity of the maser features to
take into account turbulent velocities and small deviations from the
spherical outflow model. Figs.~\ref{Fig:dipole} show the structure of
the magnetic field as well as the radial velocity structure observed
for the \water maser features. The best fitted model for a dipole
magnetic field is overplotted. While the magnetic field strengths can
be fit reasonably well, the model has a harder time simultaneously
fitting the velocity structure. Our fits show the stellar position to
be offset from the center of the \water maser emission by $\approx
70\times70$~mas to the NE. The magnetic axis of the dipole field is
pointed toward us at an inclination angle $i=40 \pm 5^\circ$, and a
position angle of $\Theta = 220 \pm 10^\circ$. The indicated errors
are the formal $1\sigma$ errors from the fit. However, as both the
magnetic field and the outflow velocity structure are likely to have
small scale structure, the exact errors are strongly model dependent
and are hard to quantify. 

The fitted values are remarkably consistent with the values found for
the magnetic field determined from OH masers, as well as with the
orientation angle determined from the \water maser distribution in M03
and \citet{M96}. We find that the magnetic field strength at the
surface of VX~Sgr corresponds to $B\approx 2.0\pm0.5$~kG. For the
$r^{-3}$ magnetic field strength dependence of the dipole field, this
is consistent with the observations of the magnetic field on the OH
masers as well as the SiO masers \citep{BM87}. In addition to a dipole
magnetic field, we also considered a solar-type magnetic field with a
magnetic field strength dependence of $r^{-2}$ on the distance to the
star. This resulted in a significantly lower likelihood.

Our ability to perform such a fit of VX~Sgr indicates that an ordered
large scale magnetic field exists. Previously, an alternative model
explained the magnetism in terms of local fields, frozen in high
density pockets. Although such small scale structure could still be
important, we conclude that it is likely that large scale magnetic
fields permeate the CSEs.

\section{Discussion}
\label{disc}

\subsection{Circular polarization of \water masers}

 The observations reveal circular polarization percentages up to $\sim
20\%$. Using the non-LTE radiative transfer method this indicates
magnetic field strengths along the maser line of sight up to $\sim
4$~G, while the fields determined using the LTE approximation are
generally $\sim 40\%$ higher. As shown in V02, the non-LTE circular
polarization spectra are typically not symmetric, while the LTE method
produces spectra with an point symmetric S-curve shape. However, due
to the data processing and the necessary removal of a scaled down
replica of the total power, it is impossible to directly observe the
non-symmetric non-LTE spectra. Still, we confirm that the non-LTE
spectra produce better fits to the observed V-spectra, as can be seen
in several of the spectra presented in
Figs.~\ref{Fig:uori},~\ref{Fig:vxsgr} and \ref{Fig:vxsgrspec}. Most of
the observed V-spectra also show narrowing that cannot be reproduced
in the LTE analysis, as was first found in V01. Similar to earlier
results \citep[e.g.][]{S79,VvL05}, we find that the \water masers are mostly
unsaturated after an analysis of the line widths and shapes of the
total intensity spectra.

We confirm in these observations again the absence of linear
polarization to a limit of $\approx 2\%$, which provides
compelling evidence against the non-Zeeman interpretation, considered
by \citet{WW98} for the SiO masers, as the cause for the observed high
circular polarization of the \water masers.  The fact that no
significant linear polarization is observed is an additional
indication that the circumstellar \water masers are unsaturated
\citep{NW91}.

\subsection{The strength and shape of the magnetic field}

\begin{figure}
%\epsscale{1.0} \plotone{bplot.ps}
   \resizebox{\hsize}{!}{\includegraphics{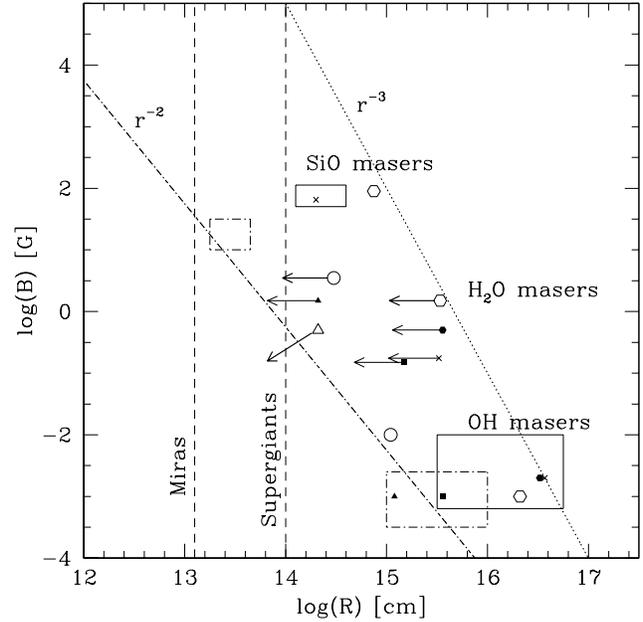}} \hfill
\caption[brplot]{Magnetic field strength, $B$, as function of
distance, $R$, from the center of the star. Dashed-dotted boxes are
the estimates for the magnetic fields in the OH and SiO maser regions
of Mira stars, solid boxes are those for the supergiants. The
dashed-dotted line indicates a solar-type magnetic field and the
dotted line indicates the dipole field. The solid symbols are the
previously observed stars (U~Her: triangles; S~Per: squares; VY~CMa:
crosses and NML~Cyg: hexagonals). The open symbols indicate the stars
of the sample presented here (U Her: triangle; VX~Sgr: hexagonal and
U~Ori: circle). Note that the open triangle only indicates the upper
limit determined for the most recent observation of U~Her. Also note
that the magnetic field strength observed on the OH masers of U~Ori is
larger than the typically observed fields for Mira variables. The
dashed lines represent estimates of the stellar radius.}
\label{Fig:brplot}
\end{figure}

 The observed magnetic field strengths on the \water masers are
consistent with the results for the other maser species assuming a $B
\propto r^\alpha$ dependence of the field strength on the distance to
the star. As the magnetic field strength determined for the SiO,
\water and OH maser features depends on the angle between the magnetic
field and the maser propagation axis, it is difficult to exactly
determine the value of $\alpha$. In Fig.~\ref{Fig:brplot} we show an
updated version of Fig.~15 from V02 including the results from the
sources observed for this paper. The open symbols indicate the stars
in the sample presented here. The points for the \water masers are
drawn at the outer radius of the maser region, which for VX~Sgr was
determined in M03. For U~Ori we used the value of $\approx 30$~AU from
\citet{BC03}. The arrows indicate a typical width of the \water maser
region. The SiO maser location of $\approx 50$~AU for VX~Sgr was taken
from \citet{DL98} and its OH maser extent ($\approx 1400$~AU) from
\citet{SC01} The OH masers around U~Ori were found to exist at
$\approx 65$~AU \citep{CC91}. For the other sources the references are
listed in the relevant section of V02. We indicate a solar-type
magnetic field ($\alpha=-2$) and a dipole field ($\alpha=-3$). The
observations of VX~Sgr indicate (as shown in \S~\ref{vxsgr}) that a
dipole field best fits the shape of the field in the \water maser
region. Extrapolating the observed magnetic field strengths to the
stellar surface, we find that Mira variable stars have surface field
strengths up to several times $10^2$~G, while supergiant stars have
fields of the order of $10^3$~G.

\subsection{The origin of the magnetic field}

 The origin of the strong, large scale magnetic fields around evolved
stars remains a topic of debate. The generation of an axisymmetric
magnetic field requires a magnetic dynamo in the interior of the
star. One of the main arguments against a dynamo generated magnetic
field is the slow rotation of the late-type star based on conservation
of angular momentum \citep{SH92}. However, models by \citet{BF01}
consider the strong differential rotation created when the degenerated
stellar core contracts while the envelope expands. In this case the
dynamo is the result of the interaction between the differential
rotation ({\bf $\omega$}) and turbulence in the convection zone around
the degenerated core ({\bf $\alpha$}). This {\bf
$\alpha\omega$}-dynamo action can produce a strong magnetic field
similar to the one observed using the circumstellar masers. It has
been argued that such strong magnetic fields should produce strong
X-ray emission. Although Mira is a weak X-ray source \citep{SK03},
observations of TX~Cam and T~Cas did not reveal any X-ray emission
\citep{KS04}. However, as the optical depth for X-ray emission around
evolved stars is expected to be high, it has been shown that this
non-detection cannot rule out the strong magnetic fields produced by
the {\bf $\alpha\omega$}-dynamo \citep{BF01}. The above models include
the interaction with a degenerated core, while the core of supergiants
such as VX~Sgr is supposedly not degenerate. However, a similar dynamo
action for supergiants, driven by the differential rotation between
the contracting non-degenerate core and the expanding outer envelope
has been shown to also be able to generate strong magnetic fields
\citep{UB82}.

 In contrast to the {\bf $\alpha\omega$}-dynamo, several papers have
examined a {\bf $\alpha^2 \omega$}-dynamo, under the assumption that
the rotation of the envelope only marginally contributes to the dynamo
action \citep{SZ02, S02}. These models produce magnetic fields that
are several orders of magnitude lower than the models using the {\bf
$\alpha\omega$}-dynamo. The observed magnetic field strength in the
maser regions of the CSE are then argued to be due to localized,
magnetized wind clumps that are produced in magnetic spots on the
surface of the star that can have local field strengths of up to $\sim
10$~G. Using another model, dynamo action of giant-cell convection at
the surface of a late-type supergiant star has recently been shown to
be able to produce fields up to $\sim 500$~G in localized spots on the
stellar surface \citep{D04}. Models that produce local strong field
however, do not readily explain the observations of large scale
magnetic field structure such as seen around VX~Sgr. Alternatively, in
the case of the {\bf $\alpha^2 \omega$}-dynamo a strong magnetic field
can be generated when the star is spun up by a close binary. The
sources in our sample do not however, show any indication of binarity,
although this cannot be ruled out.

\subsection{Shaping of the circumstellar outflow}

 The magnetic field strength of several Gauss measured in the \water
maser region implies that the magnetic pressure dominates the thermal
pressure of the circumstellar gas throughout a large part of the CSE
of both regular AGB stars and supergiants. The ratio between the
thermal and magnetic pressure is given by $\beta \equiv (8\pi n_{\rm
H} k T)/B^2)$, with $k$ the Boltzman constant. Assuming a gas density
of $n_{\rm H} = 10^9$~cm$^{-3}$ and a temperature of $T\approx10^3$~K
at the inner edge of the \water maser region, a magnetic field between
$B=0.5$ and $1$~G gives $\beta \approx 0.012 - 0.003$, indicating that
the magnetic pressure dominates the thermal pressure by factors of
$\approx 80$ or more.  In the lower density non-masing regions this
factor will be even higher unless the magnetic field is frozen into
the high density clumps. Still, in that case, when $B \propto n^k$
with $1/3 \leq k \leq 1/2$ \citep{M87}, $\beta$ will be of the
same order of magnitude.

 The effects of magnetic fields on the stellar outflow and the shaping
of the distinctly aspherical PNe have been discussed in several papers
\citep[e.g.][]{P87, CL94, G-S97}. PNe are thought to be formed due to
the interaction of the slow AGB wind ($v \sim 10$~\kms) with a
subsequent fast wind ($v \sim 1000$~\kms) generated when the central
star evolves into a white dwarf \citep{K78}. Strong, large scale,
magnetic fields around AGB stars could directly affect the fast wind
and as a result shape the PN \citep{CL94, G-S03}. Only recently,
magnetic field strengths of $1$--$3$~kG have been detected at the
surface of the central star of several PNe \citep{SWO05}. 
Another mechanism in which magnetism can play a role in shaping PNe is
described in \citet{MB00}. In the models presented in that paper, a
large scale dipole magnetic field is responsible for creating an
equatorial density enhancement in the initial, slow, AGB wind. For the
magnetic field to influence the stellar outflow, the outflow must be
sufficiently ionized, which is likely the case close to the surface of
the star. A dipole magnetic field will then only need to have
sufficient magnetic pressure ($\beta < 2.0$) to produce an equatorial
disk. The interaction of the later fast wind with this disk has been
shown to be able to easily create aspherical, cylindrical symmetric
PNe \citep[e.g.][]{I88, SL89, MEI91}. Furthermore, under the influence
of the large scale magnetic field, the circumstellar disk could become
warped \citep{Lai03}, and the interaction of the fast wind with such a
disk has been shown by \citet{I03} to be able to explain the {\it
multipolar} shape observed for several PNe, such as NGC 7027
\citep{CH02}. Although we have not directly observed the morphology of
the magnetic field for the AGB stars in our sample, the magnetic field
strengths measured on U~Ori, are fully consistent with a dipole
magnetic field such as found around the supergiant VX~Sgr. As seen in
Fig.\ref{Fig:vxsgr}, we observe indeed that the \water masers around
VX~Sgr occur in a oblate spheroid, which could be an indication of the
equatorial density enhancement expected in a CSE that has been shaped
by a dipole magnetic field.

% Even if the observed magnetic fields in the maser regions are due to
%local effects, the magnetic activity on the surface of the star that
%would be responsible for the creation of the high magnetic field maser
%clumps, can influence the stellar mass-loss and outflow (Simis \&
%Woitke 2003). Due to the magnetic activity, star-spots are formed that
%are characterized by a lower temperature than the surrounding stellar
%surface. These cool-spots will facilitate local dust formation, and,
%as a result, produce aspherical outflow. Additionally, periodic
%changes in the magnetic field, similar to the solar cycle, can enable
%periodic structure formation. This process has been suggested as the
%cause of the concentric shells that have been observed around some AGB
%stars and PNe (Soker 2000).

\section{Conclusions}
\label{concl}

 We have measured the magnetic field around the late-type stars U~Her,
U~Ori and VX~Sgr using observations of the circular polarization of
their \water masers. Although we only find an upper limit in the case
of U~Her, we find strong magnetic fields of $\sim 3.5$ and $4$~G for
U~Ori and VX~Sgr respectively. The rich structure of the \water masers
around the supergiant VX~Sgr enabled us to determine the shape of the
magnetic field around this supergiant star. The observations are best
represented by a dipole magnetic field at angles that are remarkably
consistent with those of the dipole field used to explain previous OH
maser polarization observations at much larger distances from the
central star. This confirms the presence of an ordered magnetic field
close to the star. Additionally, the ellipsoidal structure of the
\water masers around VX~Sgr is aligned with the equatorial plane of
the dipole field, which could indicate the equatorial density
enhancement caused by the magnetic field as described by
\citet{MB00}. The magnetic field strengths determined in the \water
maser regions around U~Ori and U~Her, lower mass evolved stars that
are the progenitors of PNe, are also consistent with a dipole field,
such as found around VX~Sgr. As a result, also for those stars, the
magnetic fields can cause aspherical density structures that result
in non spherically symmetric PNe.

\begin{acknowledgements}
WV acknowledges the hospitality of the Harvard-Smithsonian CfA during
his visit which was supported by the Niels Stensen Foundation.
\end{acknowledgements}


\begin{thebibliography}{99}
%\thispagestyle{headings}

%Sub-au imaging of water vapour clouds around four asymptotic giant branch stars
\bibitem[Bains et al.(2003)]{BC03} 
Bains, I., Cohen, R.~J., Louridas, A., Richards, A.~M.~S., Rosa-Gonz{\' a}lez, D., \& Yates, J.~A.\ 2003, \mnras, 342, 8 

%Evidence for strong magnetic fields in the inner envelopes of late-type stars
\bibitem[Barvainis, McIntosh, \& Predmore(1987)]{BM87} 
Barvainis, R., McIntosh, G., \& Predmore, C.~R.\ 1987, \nat, 329, 613 

%Dynamos in asymptotic-giant-branch stars as the origin of magnetic fields shaping planetary nebulae
\bibitem[Blackman et al.(2001)]{BF01} 
Blackman, E.~G., Frank, A., Markiel, J.~A., Thomas, J.~H., \& Van Horn, H.~M.\ 2001, \nat, 409, 485 

%MERLIN observations of the circumstellar envelope of VX Sagittarius
\bibitem[Chapman \& Cohen(1986)]{CC86} 
Chapman, J.~M.~\& Cohen, R.~J.\ 1986, \mnras, 220, 513 

%U Orionis - The evolution and proper motion of the OH maser envelope
\bibitem[Chapman, Cohen, \& Saikia(1991)]{CC91} 
Chapman, J.~M., Cohen, R.~J., \& Saikia, D.~J.\ 1991, \mnras, 249, 227 

%Magnetic shaping of planetary nebulae and other stellar wind bubbles
\bibitem[Chevalier \& Luo(1994)]{CL94} 
Chevalier, R.~A.~\& Luo, D.\ 1994, \apj, 421, 225 

%High resolution near-infrared spectro-imaging of NGC 7027
\bibitem[Cox et al.(2002)]{CH02} 
Cox, P., Huggins, P.~J., Maillard, J.-P., Habart, E., Morisset, C., Bachiller, R., \& Forveille, T.\ 2002, \aap, 384, 603 

%Characteristics of dust shells around 13 late-type stars
\bibitem[Danchi et al.(1994)]{DB94} 
Danchi, W.~C., Bester, M., Degiacomi, C.~G., Greenhill, L.~J., \& Townes, C.~H.\ 1994, \aj, 107, 1469 

%VLBI Imaging of the 86 GHz SiO Maser Emission in the Circumstellar Envelope of VX Sagittarii
\bibitem[Doeleman, Lonsdale, \& Greenhill(1998)]{DL98} 
Doeleman, S.~S., Lonsdale, C.~J., \& Greenhill, L.~J.\ 1998, \apj, 494, 400 

%Magnetic activity in late-type giant stars: Numerical MHD simulations of non-linear dynamo action in Betelgeuse
\bibitem[Dorch(2004)]{D04} Dorch, S.~B.~F.\ 2004, \aap, 
423, 1101 

%First polarimetric images of NML Cyg at 1612 and 1665 MHz
\bibitem[Etoka \& Diamond(2004)]{E04} 
Etoka, S.~\& Diamond, P.\ 2004, \mnras, 348, 34 

%Three-dimensional Magnetohydrodynamical Modeling of Planetary Nebulae: The Formation of Jets, Ansae, and Point-symmetric Nebulae via Magnetic Collimation
\bibitem[Garc{\'{\i}}a-Segura(1997)]{G-S97} 
Garc{\'{\i}}a-Segura, G.\ 1997, \apjl, 489, L189 

%Winds, Bubbles, ...but Magnetized: Solutions for High Speed Post-AGB Winds and Their Extreme Collimation
\bibitem[Garc{\'{\i}}a-Segura, L{\' o}pez,\& Franco(2003)]{G-S03}
Garc{\'{\i}}a-Segura, G., L{\' o}pez, J.~A., \& Franco, J.\ 2003, Revista Mexicana de Astronomia y Astrofisica Conference Series, 15, 12 

%Interferometric Observations of the SiO Masers and Dust Shell of VX Sagittarii
\bibitem[Greenhill et al.(1995)]{G95} 
Greenhill, L.~J., Colomer, F., Moran, J.~M., Backer, D.~C., Danchi, W.~C., \& Bester, M.\ 1995, \apj, 449, 365 

%%SiO maser polarization in evolved stars: magnetic field
%\bibitem[Herpin et al.(2003)]{H03} Herpin, F., Baudry, A., 
%Thum, C., Morris, D., \& Wiesemeyer, H.\ 2003, SF2A-2003: Semaine de 
%l'Astrophysique Francaise, 523

%Blowing bubbles
\bibitem[Icke(1988)]{I88} 
Icke, V.\ 1988, \aap, 202, 177 

%Blowing up warped disks
\bibitem[Icke(2003)]{I03} 
Icke, V.\ 2003, \aap, 405, L11 

%Constraining the X-Ray Luminosities of Asymptotic Giant Branch Stars: TX Camelopardalis and T Cassiopeia
\bibitem[Kastner \& Soker(2004)]{KS04} 
Kastner, J.~H.~\& Soker, N.\ 2004, \apj, 608, 978 

%Data reduction techniques for spectral line polarization VLBI observations.
\bibitem[Kemball, Diamond, \& Cotton(1995)]{KDC95} 
Kemball, A.~J., Diamond, P.~J., \& Cotton, W.~D.\ 1995, \aaps, 110, 383 

%Imaging the Magnetic Field in the Atmosphere of TX Cam
\bibitem[Kemball \& Diamond(1997)]{KD97} 
Kemball, A.~J.~\& Diamond, P.~J.\ 1997, \apjl, 481, L111 

%Multiple Molecular Winds in Evolved Stars. I. A Survey of CO (2-1) and CO (3-2) Emission from 45 Nearby AGB Stars
\bibitem[Knapp et al.(1998)]{KY98} 
Knapp, G.~R., Young, K., Lee, E., \& Jorissen, A.\ 1998, \apjs, 117, 209 

%On the origin of planetary nebulae
\bibitem[Kwok, Purton, \& Fitzgerald(1978)]{K78} 
Kwok, S., Purton, C.~R., \& Fitzgerald, P.~M.\ 1978, \apjl, 219, L125 

%The Spatial Structure of Silicon Monoxide Masers
\bibitem[Lane(1984)]{L84} 
Lane, A.~P.\ 1984, IAU Symp.~110: VLBI and Compact Radio Sources, 110, 329 

%Warping of Accretion Disks with Magnetically Driven Outflows: A Possible Origin for Jet Precession
\bibitem[Lai(2003)]{Lai03} 
Lai, D.\ 2003, \apjl, 591, L119 

%VLBI Observations of OH stars: S Persei
\bibitem[Masheder et al.(1999)]{MvL99} 
Masheder, M.~R.~W., van Langevelde, H.~J., Richards, A.~M.~S., Greenhill, L., \& Gray, M.~D.\ 1999, New Astronomy Review, 43, 563 

%Disk Formation by Asymptotic Giant Branch Winds in Dipole Magnetic Fields
\bibitem[Matt et al.(2000)]{MB00} 
Matt, S., Balick, B., Winglee, R., \& Goodson, A.\ 2000, \apj, 545, 965 

%The Circumstellar Enivronment of Evolved Stars as Revealed by Studies of Circumstellar Water Masers
\bibitem[Marvel(1996)]{M96} 
Marvel, K.~B.\ 1996, Ph.D.~Thesis, New Mexico State Univ.

%Hydrodynamical models of aspherical planetary nebulae
\bibitem[Mellema, Eulderink, \& Icke(1991)]{MEI91} 
Mellema, G., Eulderink, F., \& Icke, V.\ 1991, \aap, 252, 718 

%Star formation in magnetic interstellar clouds. I - Interplay between theory and observations. II - Basic theory
\bibitem[Mouschovias(1987)]{M87} 
Mouschovias, T.~C.\ 1987, NATO ASIC Proc.~210: Physical Processes in Interstellar Clouds, 453 

%The radially expanding molecular outflow of VX Sagittarii
\bibitem[Murakawa et al.(2003)]{MYRC03} 
Murakawa, K., Yates, J.~A., Richards, A.~M.~S., \& Cohen, R.~J.\ 2003, \mnras, 344, 1 (M03)

%Spectral line profiles and luminosities of astrophysical water masers
\bibitem[Nedoluha \& Watson(1991)]{NW91} 
Nedoluha, G.~E.~\& Watson, W.~D.\ 1991, \apjl, 367, L63 

%The Zeeman effect in astrophysical water masers and the observation of strong magnetic fields in regions of star formation
\bibitem[Nedoluha \& Watson(1992)]{NW92} 
Nedoluha, G.~E.~\& Watson, W.~D.\ 1992, \apj, 384, 185 

%Models of OH Maser Variations in U Herculis
\bibitem[Palen \& Fix(2000)]{PF00} 
Palen, S.~\& Fix, J.~D.\ 2000, \apj, 531, 391 

%The nature of bipolar planetary nebulae
\bibitem[Pascoli(1987)]{P87} 
Pascoli, G.\ 1987, \aap, 180, 191 

%Stellar OH masers and magnetic fields - VLBI observations of U Orionis and IRC +10420
\bibitem[Reid et al.(1979)]{R79} 
Reid, M.~J., Moran, J.~M., Leach, R.~W., Ball, J.~A., Johnston, K.~J., Spencer, J.~H., \& Swenson, G.~W.\ 1979, \apjl, 227, L89 

%OH masers and the structure of Mira/Red Supergiant winds
\bibitem[Richards et al.(2000)]{RC00} 
Richards, A.~M.~S., Cohen, R.~J., Murakawa, K., Yates, J.~A., van Langevelde, H.~J., Gray, M.~D., Masheder, M.~R.~W., \& Szymczak, M.~D.\ 2000, Proc. of the 5th EVN Symp. , Eds.: J.E.~Conway et al., publ. Onsala Space Observatory, p.~185, 185 

%OH masers and the structure of Mira/Red Supergiant winds
%\bibitem[Richards et al.(2000)]{RC00} 
%Richards, A.~M.~S., Cohen, R.~J., Murakawa, K., Yates, J.~A., van Langevelde, H.~J., Gray, M.~D., Masheder, M.~R.~W., \& Szymczak, M.~D.\ 2000, EVN Symposium 2000, Proceedings of the 5th european VLBI Network Symposium held at Chalmers University of Technology, Gothenburg, Sweden, June 29 - July 1, 2000, Eds.: J.E.~Conway, A.G.~Polatidis, R.S.~Booth and Y.M.~Pihlstr{\" o}m, published Onsala Space Observatory, p.~185, 185 

%Interacting winds and the shaping of planetary nebulae
\bibitem[Soker \& Livio(1989)]{SL89} 
Soker, N.~\& Livio, M.\ 1989, \apj, 339, 268 

%Can a single AGB star form an axially symmetric planetary nebula?
\bibitem[Soker \& Harpaz(1992)]{SH92} 
Soker, N.~\& Harpaz, A.\ 1992, \pasp, 104, 923 

%Turbulent dynamo in asymptotic giant branch stars
\bibitem[Soker \& Zoabi(2002)]{SZ02} 
Soker, N.~\& Zoabi, E.\ 2002, \mnras, 329, 204 

%Local circumstellar magnetic fields around evolved stars
\bibitem[Soker(2002)]{S02} 
Soker, N.\ 2002, \mnras, 336, 826 

%Magnetic Flares on Asymptotic Giant Branch Stars
\bibitem[Soker \& Kastner(2003)]{SK03} 
Soker, N.~\& Kastner, J.~H.\ 2003, \apj, 592, 498 

%The structure of H2O masers associated with late-type stars
\bibitem[Spencer et al.(1979)]{S79} 
Spencer, J.~H., Johnston, K.~J., Moran, J.~M., Reid, M.~J., \& Walker, R.~C.\ 1979, \apj, 230, 449 

%Discovery of magnetic fields in central stars of PNe
\bibitem[Jordan, Werner \& O'Toole(2005)]{SWO05}
Jordan, S., Werner, K.~\& O'Toole, S.~J.\ 2005, \aap, accepted

%Polarization structure of the OH 1612-MHz maser envelope of VXSagittarii
\bibitem[Szymczak \& Cohen(1997)]{SC97} 
Szymczak, M.~\& Cohen, R.~J.\ 1997, \mnras, 288, 945 

%Magnetic field structure in the outer OH maser envelope of VX Sagittarii
\bibitem[Szymczak, Cohen, \& Richards(2001)]{SC01} 
Szymczak, M., Cohen, R.~J., \& Richards, A.~M.~S.\ 2001, \aap, 371, 1012 

%The circumstellar magnetic field of VX Sagittarii
\bibitem[Trigilio, Umana, \& Cohen(1998)]{T98} 
Trigilio, C., Umana, G., \& Cohen, R.~J.\ 1998, \mnras, 297, 497 

%Chromospheric activity of late-type giants and supergiants Reappearance of dynamo activity in the interior due to the spin-up of the core in evolution
\bibitem[Uchida \& Bappu(1982)]{UB82} 
Uchida, Y.~\& Bappu, M.~K.~V.\ 1982, Journal of Astrophysics and Astronomy, 3, 277 

%Circular polarization of circumstellar water masers around S Per
\bibitem[Vlemmings, Diamond, \& van Langevelde(2001)]{V01} 
Vlemmings, W., Diamond, P.~J., \& van Langevelde, H.~J.\ 2001, \aap, 375, L1 (V01)

%Astrometry of the stellar image of U Her amplified by the circumstellar 22 GHz water masers
\bibitem[Vlemmings, van Langevelde, \& Diamond(2002)]{V02a} 
Vlemmings, W.~H.~T., van Langevelde, H.~J., \& Diamond, P.~J.\ 2002, \aap, 393, L33 

%Circular polarization of water masers in the circumstellar envelopes of late type stars
\bibitem[Vlemmings, Diamond, \& van Langevelde(2002)]{V02b} 
Vlemmings, W.~H.~T., Diamond, P.~J., \& van Langevelde, H.~J.\ 2002, \aap, 394, 589 (V02)

%VLBI astrometry of circumstellar OH masers: Proper motions and parallaxes of four AGB stars
\bibitem[Vlemmings et al.(2003)]{V03} 
Vlemmings, W.~H.~T., van Langevelde, H.~J., Diamond, P.~J., Habing, H.~J., \& Schilizzi, R.~T.\ 2003, \aap, 407, 213 

%Astrophysical water masers: line profiles analysis
\bibitem[Vlemmings \& van~Langevelde.(2005)]{VvL05} 
Vlemmings, W.~H.~T., van Langevelde, H.~J., 2005, astro-ph/0501627

%A Non-Zeeman Interpretation for Polarized Maser Radiation and the Magnetic Field at the Atmospheres of Late-Type Giants
\bibitem[Wiebe \& Watson(1998)]{WW98} 
Wiebe, D.~S.~\& Watson, W.~D.\ 1998, \apjl, 503, L71 

%The Spatial Distribution of Circularly Polarized 1612 MHz OH Maser Emission From VX SGR
\bibitem[Zell \& Fix(1996)]{ZF96} 
Zell, P.~J.~\& Fix, J.~D.\ 1996, \aj, 112, 252 

\end{thebibliography}
\end{document}